# Which international co-authorships produce higher quality journal articles?[1]

Mike Thelwall: Information School, University of Sheffield, UK, and Statistical Cybermetrics and Research Evaluation Group, University of Wolverhampton, UK.
Kayvan Kousha: Statistical Cybermetrics and Research Evaluation Group, University of Wolverhampton, UK.
Mahshid Abdoli: Statistical Cybermetrics and Research Evaluation Group, University of Wolverhampton, UK.
Emma Stuart: Statistical Cybermetrics and Research Evaluation Group, University of Wolverhampton, UK.
Meiko Makita: School of Health Sciences, University of Dundee, UK, and Statistical Cybermetrics and Research Evaluation Group, University of Wolverhampton, UK.
Paul Wilson: Statistical Cybermetrics and Research Evaluation Group, University of Wolverhampton, UK.
Jonathan Levitt: Statistical Cybermetrics and Research Evaluation Group, University of Wolverhampton, UK.

International collaboration is sometimes encouraged in the belief that it generates higher quality research or is more capable of addressing societal problems. Nevertheless, whilst there is evidence that the journal articles of international teams tend to be more cited than average, perhaps from increased international audiences, there is no science-wide direct academic evidence of a connection between international collaboration and research quality. This article empirically investigates the connection between international collaboration and research quality for the first time, with 148,977 UK-based journal articles with post publication expert review scores from the 2021 Research Excellence Framework (REF). Using an ordinal regression model controlling for collaboration, international partners increased the odds of higher quality scores in 27 out of 34 Units of Assessment (UoAs) and all Main Panels. The results therefore give the first large scale evidence of the fields in which international co-authorship for articles is usually beneficial. At the country level, the results suggests that UK collaboration with other high research-expenditure economies generates higher quality research, even when the countries produce lower citation impact journal articles than the UK. Worryingly, collaborations with lower research-expenditure economies tend to be judged lower quality, possibly through misunderstanding Global South research goals.
**Keywords**: Research collaboration; co-authorship; scientometrics; research policy; REF2021; Research assessment; research quality

## 1 Introduction

International collaboration has grown consistently over the last century (Larivière et al., 2015; Hsiehchen et al., 2018). Although this is partly due to faster communication, cheaper travel, and international mobility/migration (Adams et al., 2005; Freeman et al., 2014; Kato & Ando, 2017), it is widely encouraged by research funders in the belief that it is beneficial for scientific progress (Olechnicka et al., 2019). This is a contentious issue because many national research funding programs primarily or only finance local researchers but may also have grants specifically for international collaboration. This creates a tension between encouraging international collaboration for possible higher quality research or networking benefits and partly avoiding it to focus on local benefits and national researchers.

---
[1] Thelwall, M., Kousha, K., Abdoli, M., Stuart, E., Makita, M., Wilson, P. & Levitt, J. (in press). Which international co-authorships produce higher quality journal articles? *J. Association for Information Science & Technology*.

The strategy of promoting international collaboration is indirectly supported by evidence of the greater citation impact of internationally co-authored research (Zhou et al., 2020). This is in turn supported by evidence that more cited research tends to be higher quality, especially in the physical and health sciences (Thelwall et al., 2023c). Nevertheless, this citation advantage is at least partly an audience effect: more people aware of the work and citing it due to multiple national networks recognising the authors (Thelwall et al., 2023b; Wagner et al., 2019). Moreover, research quality is more fundamental and important for assessing research than citation counts. The latter are weak proxies for research quality because citations primarily reflect academic significance rather than societal value or the two other recognised dimensions of research quality: rigour and originality (Langfeldt et al., 2020). In parallel, whilst it is known that research with more authors tends to be higher quality, except in the arts and humanities (at least for the UK: Thelwall et al., 2023b; see also: Presser, 1980), previous research has rarely separated international authorship from collaboration in general, so it is not known whether internationalism conveys a quality advantage above that from general collaboration. Moreover, citation-based studies suggest that not all international collaboration is equal because some countries benefit more (Zhou et al., 2020) and it is unclear whether this is also true for research quality.

Only two previous studies have directly investigated the relationship between research quality (rather than citation counts) and author team internationalism, factoring out collaboration. A regression analysis of 16,554 biomedical articles published before 2013 with at least two post-publication peer review quality scores from the F1000Prime website found both the number of countries and the number of authors to be significant predictors of higher quality scores (Bornmann, 2017), but the results are limited to a single field. A white paper analysing all academic disciplines used a stepwise (binary) logistic regression analysis of UK journal articles 2008-13 to develop a model to predict whether an article would receive the highest quality score given by post-publication expert review (with similar review procedures to that used in the data used for the current article, as described below). It found author team internationality (as a binary variable) to be a significant predictor in only three out of 36 fields examined (Clinical Medicine; Earth Systems and Environmental Sciences; Computer Science and Informatics) (HEFCE, 2015). The regression included 32 independent variables, many partly overlapping, which undermined its power to detect the relationship between international collaboration for research quality. In particular, including citation counts and the number of countries represented in an authorship team as independent variables make its results uninformative about the role of international collaboration. Moreover, neither of these two studies differentiated between the countries involved in the collaboration.

In response to the almost complete lack of evidence of a relationship between research quality (rather than citation counts) and international collaboration, with the partial exceptions of the weak effects in the two contexts noted above, this article investigates the issue for all fields of science and with the largest and most recent dataset yet. It focuses on the UK for the pragmatic reason of data availability. In addition, this study investigates for the first time whether there are differences between countries in their apparent contributions to the quality (rather than citation counts) of internationally co-authored research. Here, the quality scores used are taken from the UK Research Excellence Framework (REF), as assigned by expert reviewers on a four-point scale. The following research questions drive the study.
1. In which fields are UK-authored journal articles higher quality when they are internationally co-authored?
2. Does the quality of an internationally collaborative journal article with the UK depend on which countries collaborate?
3. Are high impact international partners for the UK necessary for international journal articles to be higher quality?

# 2 Background and hypotheses

## 2.1 Field differences in the advantages of international teams

International collaboration is not uniform across fields. In professional contexts, including some aspects of healthcare, nationally organised services are obstacles to international collaboration because working practices differ (e.g., Gladstone et al., 2019). Law is perhaps an extreme example, with each country having different legal frameworks. This may be partly why international research collaboration is greater in the natural sciences than the social sciences, and rare in the humanities (Lariviere et al., 2006; Gazni et al., 2012).

Although previous research has not investigated field differences in the relationship between international teams and research quality (with the partial HEFCE exception mentioned above), there are known to be field differences in the citation advantage of international collaboration, as suggested by descriptive studies analysing various citation impact and collaboration indicators. One early study found that international research tended to be more cited in all six fields examined: Clinical medicine, Biomedical research, Biology, Chemistry, Physics, Mathematics, Engineering and Earth and space sciences. Nevertheless, in some country/field combinations, international collaborative research was less cited than domestic research (Glänzel, 2001), so the pattern is not universal. A later detailed analysis of international differences in the citation advantage of international collaboration in 2004 for 60 countries found the greatest advantages for social science and engineering journal articles. The international collaboration advantage for physics and medicine was lower but still substantial (Lancho-Barrantes et al., 2013). It is not clear why international collaboration should be most advantageous in social sciences and engineering, however. Whitley's (2000) theory of organisational dimensions of disciplinary differences suggests that impact in fields with little agreement on what is worth researching (i.e., high technical task uncertainty) might be harder to gain and so international networks might help with wider audiences. This seems likely to include most social sciences, but not engineering. For engineering (and many professional fields), international research might be more likely to focus on general problems rather than local solutions. Despite the lack of clear theoretical motivations, H1 is based on the best available empirical evidence (Lancho-Barrantes et al., 2013).
- **H1**: UK international collaboration is most advantageous in the social sciences and engineering.

## 2.2 National differences in the advantages of international teams

In addition to the evidence that international collaboration sometimes associates with higher research quality from the two studies mentioned above (Bornmann, 2017; HEFCE, 2015) there is substantial evidence that international collaboration in general associates with higher citation rates (Hall et al., 2018; Larivière et al., 2015; Zhou et al., 2020) and almost all countries tend to attract more citations per paper for their international collaborations overall (Adams, 2013; Hsiehchen et al., 2018). It may not increase productivity, however (e.g., Abramo et al., 2017). Since research collaboration of any type usually associates with more citations (Thelwall & Maflahi, 2020) and higher quality (Thelwall et al., 2023b), it is important to distinguish between the general benefits of collaboration and those that are specific to international partnerships. The remainder of this subsection covers citation analysis rather than quality analysis due to a lack of relevant research quality studies.

There are national differences in the citation advantage of international collaboration. The country with the lowest citation impact in a collaboration tends to gain the most from it (Lancho-Barrantes et al., 2013; Guerrero Bote et al., 2013), except for collaboration with the USA (which has a unique bibliometric advantage in citation databases: Gingras & Khelfaoui, 2018). Collaboration with Iran may also not be beneficial overall, although it is not clear why (Guerrero Bote et al., 2013). Less resourced countries seem to benefit most (in terms of citations) from international collaboration with higher resourced countries when the latter leads it (Chinchilla-Rodríguez et al., 2019). Ironically, this type of research seems likely to be disproportionately neo-colonial or helicopter science, taking place in less developed countries and with quality and impact suffering from insufficient local involvement (Minasny et al., 2020). A regression-based analysis of the number of citations or Mendeley readers for biochemistry research articles found that having a US, British, German, French or Canadian author was an advantage, but an Indian and Italian (readers only) author was a disadvantage (Sud & Thelwall, 2016). Differences in the citation rates of international collaborations may also be affected by the willingness of national funding bodies to include overseas partners in projects primarily targeting local researchers. In this context, the USA and China seem particularly willing to fund overseas research collaborators (Huang & Huang, 2018). The strongest evidence leads to the second research hypothesis.

- **H2**: UK International collaboration is most advantageous with high citation countries.

National concentrations of resources may also generate specific field-country combinations in international collaboration advantages. For example, the presence of the European Organization for Nuclear Research (CERN) in Switzerland may generate natural citation advantages for all non-Swiss particle physicists when they need to access expensive equipment. Similarly, the presence of MagLab in Florida might help to generate an overall citation advantage for magnetic field researchers collaborating with the USA if MagLab collaborations tend to be more cited.

From a Global South perspective, the advantages of collaboration with Global North countries might depend on whether their concerns and needs are properly incorporated into the process (Chinchilla-Rodríguez et al., 2018). It is important to avoid citation counting when evaluating Global South research because this foregrounds Global North priorities (Barrere, 2020) in addition to being affected by citation database biases.

## 2.3 *Research funding and the advantages of international teams*

International research might tend to be higher quality if it is better funded. This is because funded research tends to be higher quality (Thelwall et al., 2023a), some funding streams require international collaboration (e.g., most EU programmes), and some large-scale projects need to be international to attract sufficient funding for expensive equipment (Thelwall, 2020). Even national funding streams can include the option to add international partners (Huang & Huang, 2018; Zhou et al., 2020) and this may improve success rates, even if full funding of international partners in national projects is not supported. From a different perspective, researchers may believe that international collaboration is necessary to attract funding (Kwiek, 2021). All these factors would increase the chance that international collaborations are funded, although there is no systematic direct overall evidence of this. In medicine, international studies can increase the size of a dataset either by combining resources for labour intensive, expensive data collection or for access to a greater number of subjects with a particular condition (Cooley et al., 2003). This can produce more definitive, and presumably more significant, results, but would need more funding than more limited national research. Indirectly supporting the funding hypothesis, better funded US researchers have been shown to have wider collaboration networks (Bozeman & Corley, 2004).

A confounding factor is that "better" researchers might be more successful at attracting large international grants (e.g., most EU funding) or international collaborators, so the primary advantage may be the researchers able to get the funding rather than the value of the grant. For example, highly cited researchers might be better positioned to reach out to overseas researchers to gain access to expensive equipment (e.g., Beaver, 2001), complementary skills or high-level expertise. National funding programs may also aim to connect local excellent researchers with high quality experts from other countries (Bloch et al., 2015; Edler, 2012). There is empirical evidence from two contexts (Italy and chemistry) to suggest that the overall citation advantage of international research is at least partly from more highly cited researchers engaging in it (Abramo et al., 2011; Kato & Ando, 2013). For chemistry research, the citation level of the participating researchers is not enough to explain the citation rates of their international collaborations, however (Kato & Ando, 2013).

The amount of money spent directly on a research study is not systematically recorded in publications, although the name of the funder often is. In addition, the indirect spending on research, in terms of equipment and resources for the training and routine research of the academics involved is rarely, if ever calculated. In this context, it seems reasonable to assume that scholars in countries spending more on research would tend to be able to draw directly and indirectly on greater financial resources for their work.

- **H3**: UK international collaboration is more advantageous when the international collaborators are from countries that spend more on research than when the international collaborators are from countries that spend less on research.

## 2.4 Other reasons for the advantages of international co-authorship

International research may produce higher quality research for reasons that are difficult to systematically quantify. For example, a medical article might generate substantial value by combining patient data from multiple countries to identify statistically significant conclusions that were not possible with previous single-country studies (e.g., Bolton et al., 2021; Tritschler et al., 2020), or may generate larger sample sizes for the same cost by recruiting in lower income countries (Thiers et al., 2008). The latter may sometimes be helicopter science, however (Schutte & Karim, 2023).

Diversity in teams has been shown to be advantageous in some contexts (Curşeu & Pluut, 2013; Joshi & Roh, 2009), so it is plausible that an international team of researchers might benefit from greater variety in ideas, skills, expertise, resources, scientific cultures, and background knowledge (Wagner et al., 2019). For social science research, international collaboration can give access to different social and professional contexts for comparative or more diverse studies, as well as the expertise of the researchers involved (Teune, 1966). An analysis of Web of Science journal articles and conference papers 1973-2009 found a tendency for countries to collaborate with other countries having dissimilar levels of citations per paper (Hsiehchen et al., 2018). Nevertheless, the empirical evidence for the benefits of diversity in academia is not extensive (Hall et al., 2018) and international research seems to be more conventional (Wagner et al., 2019), with the latter point partly undermining the diversity claim. In fields where there is a high degree of agreement on tasks and goals (i.e., technical task certainty: Whitley, 2000), such as medicine, novelty may be less important than other aspects of quality, and obtaining adequate funding may be of primary importance (Gottlieb et al., 2019).

### 2.5  Other benefits of international collaboration

International collaboration may be promoted for reasons other than its perceived direct research quality benefits. It may be regarded as developmentally positive for the researchers involved, helping them to build networks and experiences that will benefit their later careers (Corley et al., 2019; see also Dusdal & Powell, 2021). Overlapping with citation impact considerations, it may be necessary for some types of research (irrespective of its quality) that is too expensive for individual countries or when there is too little national data to obtain strong result about a medical or other issue. International bodies, such as the European Union, may also want to promote international collaboration as part of a wider policy of integration, such as the European Research Area (Defazio et al., 2009; Georghiou, 2001; Kwiek, 2021).

For researchers in the Global South, engaging with academics and priorities from the Global North may be a route to get international recognition for their work (Martinez & Sá, 2020). This might provide an incentive to collaborate with Global North researchers, partly explaining the dissimilar country collaboration tendency. Conversely, others may avoid collaborating with the Global North to avoid neo-colonialist research practices, irrelevant research priorities and counterproductive quality assessment practices. Such researchers would not appear in the REF data analysed here.

## 3  Methods

There are many types of research collaboration, not all of which lead to co-authored journal articles (Katz & Martin, 1997), but this study focuses on journal articles having authors affiliated with at least two countries (including the UK for pragmatic reasons). A regression approach was used to model the effects of national contributions to the quality of journal articles, using a large multidisciplinary set of quality scores for individual UK journal articles.

### 3.1  Data

The raw data consisted of 148,977 journal articles submitted to the national research evaluation exercise REF2021 by UK universities and first published between 2014 and 2020. Each UK academic with a research component of their job was allowed to submit up to 5 outputs, with each institution submitting an average of 2.5 outputs per full-time equivalent research active member of staff and these outputs were presumably considered to be the submitting researchers' best work. Members of staff and their associated outputs were submitted to one of 34 Units of Assessment (UoAs), each of which encompasses a broad area of research (see the first graph below for a list of names). Whilst outputs in low numbered UoAs (health and physical sciences) tended to be journal articles, in higher numbered UoAs (arts and humanities) they were more likely to be books, book chapters, performances, or artworks. The assessment was UK-wide and simultaneous for all fields and institutions.

The Research Excellence Framework articles had been assessed by at least two peer reviewers with the most relevant expertise from within the UoA subpanel team (as decided by the UoA chair). These were all field experts and usually senior UK academics [full professors] from a preselected set of over 1000 that is published online (REF, 2021). Unlike the Italian equivalent (Galderisi et al., 2019), external experts are not consulted for the reviewing. The assessors assigned a single overall quality score: 1* nationally recognised, 2* internationally recognised, 3* internationally excellent or 4* world leading. Quality was judged for originality, significance, and rigour, with detailed instructions to guide assessors for the four different quality levels (p. 34-41 of: REF, 2019).

None of the criteria for any panel or overall mention international collaboration, although for UoAs 1-6, assessors can consider, "the scale, challenge and logistical difficulty posed by the research" (p. 35 of: REF, 2019), which may be related to the extent of collaboration involved. Since the core aspects of research quality were considered to have disciplinary differences, different instructions were provided for each of the four Main Panels (see below for context about these). This is an expert post-publication single-blind peer review exercise. Assessors could not score or discuss outputs where there was a perceived conflict of interest (p. 98-100 of: REF, 2019).

There were extensive procedures to ensure scoring consistency within UoAs and additional cross-UoA steps. All assessors were selected by an open call and were provided with the public guidelines and trained in the key procedures and requirements. Within UoAs, outputs were independently assessed and scored by two assessors (sometimes three, but exact numbers are unknown), who then met to agree a joint score. The scores were then discussed with the rest of the assessors for the UoA, checking for consistency and lenient/generous scoring/scorers. There were also group discussions with each UoA and across UoAs to discuss the scoring process and related issues. There were REF-wide statistical checks for bias based on author gender, interdisciplinarity and early career researcher status. Since the results direct 100% of the UK block grant for research, approximately £14 billion over seven years, all stages of the process are likely to have been taken seriously.

Despite the above careful procedures, the REF definition of research quality is UK-based and specific to the purposes of the REF. Field-based concepts of research quality may differ, for example, by emphasising disciplinary goals or by considering other factors. In addition, Global South research goals may foreground factors of importance to the Global Majority, such as sustainable development, equity, and knowledge with practical value (e.g., Grieve & Mitchell, 2020; Mahajan et al., 2023; Miguel et al., 2023). In particular, local societal context and benefits together with local capacity building are more important in the Global South (Lebel & McLean, 2018; Kraemer-Mbula et al., 2020).

In 11 UoAs, assessors could consult a table of Web of Science citation percentile thresholds provided by the REF team when they couldn't resolve disagreements about article quality scores but other than this the assessment process was pure peer review. Moreover, the assessors were repeatedly instructed to ignore journal impact factors and journal rankings (and the reputation of the publishing journal) as well as other sources of citation counts.

The 318 articles with a score of 0 were discarded for being potentially ineligible for reasons unrelated to quality (e.g., the submitting author was judged not to have made a significant contribution, or the article was judged to be a review). Articles without UK authors were also discarded, these potentially being for authors that had moved to a UK institution during the REF period. This study only includes the journal articles and their provisional

REF2021 scores from March 2021, as supplied confidentially by the REF team. The raw data was deleted, as required by the REF team, in May 2021.

A journal article was sometimes submitted by different co-authors to the same or a different UoA. Such duplicates were removed and, in cases where an article had received different scores, the median score was used, or a random selection when there were two medians. The UoAs are grouped thematically into the four REF Main Panels A (mainly health and life sciences): UoAs 1-6; B (mainly physical sciences and engineering): UoAs 7-12; C (mainly social sciences): UoAs 13-24; D (mainly arts and humanities): UoAs 25-34 for some analyses to give additional statistical power to analyse country-level differences. When grouped into main panels, duplicate articles were again eliminated using the same procedure as before for differing quality scores.

### 3.2 RQ1, H1: Regression Analysis for internationalism in UK research authorship

The value of an international dimension in collaboration was modelled using ordinal linear regression (Harrell, 2015) applied to a binary variable for internationalism (at least one non-UK author or not), with the log of the number of authors as a control variable. The model therefore tests whether internationality rather than general collaboration associates with higher quality research. The model treats all international collaborators the same and all collaboration proportions the same. The regression equation is as follows, where α is the constant term, *international* is a binary variable and *authors* is the number of authors. Changing *international* to the proportion of non-UK contributions does not change the overall results substantially. The primary variable of interest is $\beta_i$ but $\beta_a$ is also reported.

$$score = \alpha + \beta_i international + \beta_a \log(authors) \qquad (1)$$

The team size logarithm term helps to account for larger teams tending to generate work with higher research quality. A logarithm term is a reasonable approximation to the pattern that the citation increment of additional authors decreases as the team size increases (e.g., Thelwall & Maflahi, 2020). Thus, the model factors out the collaboration quality bonus from any international bonus.

Ordinal logistic regression was used because the scores 1*, 2*, 3*, and 4* are ordered but not necessarily on the numerical scale 1-2-3-4, so applying linear regression would require an additional assumption about the distance apart of the four scores. Financially, 1* and 2* are equivalent (unfunded) and a 4* score attracts four times as much money as a 3* score, so a linear assumption would not fit with how the data is used for research funding. Ordinal logistic regression treats the four scores as ordered but not necessarily equidistant. The main assumption of the ordinal logistic regression approach used is that factors affecting the odds of an article attaining a higher score rather than a lower score do not depend on where the cut-off is set. Whilst it is not known whether this is true, it is a necessary assumption to run this kind of regression and it seems to be a reasonable hypothesis, in the absence of evidence to the contrary.

Ordinal logistic regression is similar to logistic regression except that the dependant variable is an ordered set. As in logistic regression, the regression coefficients can be transformed into odds ratios. An exponentiated regression coefficient (as shown in the graphs) of 1 indicates that the specified variable (e.g., at least one author from a specified country, logged number of authors) has no effect on the probability of an article attaining any particular REF score. Values above 1 increase the odds that an article gets a higher score and values below 1 decrease these odds.

Separate regressions were conducted for each UoA (n=34) and Main Panel (n=4) because the literature review suggests that there may be substantial differences between fields in the effects of all collaboration and of international collaboration.

### *3.3   RQ2, H1: Regression Analysis for country contributions to partly UK papers*

Apparent national author contributions to REF scores were modelled again using ordinal logistic regression controlling for author numbers but this time applied to fractional author contributions by country. Thus, the expected quality score of an article is estimated by the regression from the percentage contributions of each nation's co-authors. This allows, for example, a co-authorship with country X to reduce the probability of a high-quality article but a co-authorship with country Y to increase it relative to a UK-only article. The regression coefficients therefore estimate the relative quality contributions of each country's authors, when collaborating with the UK. The natural log of the number of authors was again added to control for general collaboration advantages to separate them from international collaboration. The regression equation is as follows, where *country* is the fractional contribution of a country to the paper (0 if not a co-author). The UK is not included because it is a redundant variable, so the constant term α represents an article with one UK author. One of the countries is "Other", combining countries that do not meet the authorship threshold for inclusion (see figure captions).

$$score = \alpha + \sum \beta_{country} country + \beta_a \log(authors) \qquad (2)$$

The underlying model is that the quality of an article is contributed to equally by all authors, with this contribution potentially varying by affiliation country share. For example, an article with seven UK authors and three Kenyan authors would have 70% UK and 30% Kenyan contributions. According to this model, the probability that this article has any given REF star rating depends on these two inputs, as well as the (natural) log of the total number of authors: log(10) in this case. The country component of this model is a simplification because the first author probably had the most responsibility for the article, or the corresponding author in some fields (e.g., as believed in medicine: Bhandari et al., 2014), and the last author may have had more responsibility than others (Duffy, 2017). The main evidence so far suggests that the first author tends to be the most important contributor to an article (Larivière et al., 2016). Nevertheless, there is no agreed weighting scheme for author contributions and non-standard author ordering, such as alphabetical and partial alphabetical, are also common (Mongeon et al., 2017). There did not seem to be a strong alternative to the equal contribution hypothesis because of these factors. Separate regressions were conducted for each Main Panel (n=4) but not for individual UoAs due to relatively little data for many countries.

It would also have been possible to use binary variables for country contributions but does not seem reasonable for highly collaborative articles and would penalise the regression coefficients of countries frequently involved in them. Although the fractional author counting by country takes into account country contributions, if any country's UK co-authorships were primarily small roles in massively multi-authored papers, this could translate into an apparent country advantage (e.g., Guerrero Bote et al., 2013). The team size factor included reduces the chance that this happens.

## 3.4 RQ3, H2, H3: Correlation analyses for factors associating with country differences in UK collaboration advantages

The RQ2 research quality ordinal regression coefficients of countries were correlated against national citation impact and funding indicators to look for patterns in the types of countries benefit UK collaborations the most, addressing RQ3 and testing H2 and H3. The average citation impact of the collaborating country was used for RQ3 and H2, as in previous investigations into the citation advantage of international collaboration (Lancho-Barrantes et al., 2013; Guerrero Bote et al., 2013).

Average country citation impact was calculated using a copy of Scopus downloaded in January-February 2022. The Mean Normalised Log-transformed Citation Score (MNLCS) was calculated for each country (Thelwall, 2017). This uses field and year normalisation for comparability between fields, so that a value of 1 always indicates world average citation rate, irrespective of field and year. All citation counts were log-transformed with ln(1+x) in the calculation to reduce the influence of individual highly cited articles. Departing from standard MLNCS practice but in keeping with the regression approach used (see below), country MNLCS were calculated on a fractional counting basis, so an article with 70% Kenyan authors would count as 0.7 of an article from Kenya. For each country the formula is:

$$MNLCS = (\sum_{i=1}^{n} \text{fraction}_i \times \ln(1 + citations_i) / field_i)/(\sum_{i=1}^{n} \text{fraction}_i) \quad (3)$$

Where $fraction_i$ is the fractional contribution of the country to article *i*, and $field_i$ is the average of the log normalised citations of all articles published in the same field and year as article *i*, irrespective of the authors' countries. We used country thresholds of a minimum 10, 25, 50 and 100 articles (fractional counting) and the results were similar for the overlapping countries so only the threshold 10 results are reported. Countries not meeting the threshold were assigned to an "Other" category in each case.

An indicator of research funding was sought to address H3. Per capita Gross Domestic Product (GDP) was considered as an indicator of national wealth since this may influence funding (it correlates positively with per capita GDP: Satish, 2021) and the amount (Clarke et al., 2007), citation rate (strong correlation: Satish, 2021) or quality (Winnik et al., 2013) of research. The United Nations Development Programme Human Development Index (HDI) and its education component (hdr.undp.org/en/data) were also considered as relevant factors (e.g., Parreira et al., 2017). These reflect overall human development and its educational component, both of which may be relevant to the capability of a nation's researchers. Nevertheless, the most appropriate available research funding indicator seems to be Gross domestic Expenditure on Research and Development (GERD) as a percentage of GDP (http://data.uis.unesco.org/Index.aspx?DataSetCode=SCN_DS&lang=en) even though it includes development funding and is not restricted to academic sector research. GERD encapsulates, "the total expenditure (current and capital) on R&D carried out by all resident companies, research institutes, university and government laboratories, etc., in a country" (https://data.oecd.org/rd/gross-domestic-spending-on-r-d.htm). GERD is also a national research capacity indicator (Lancho-Barrantes et al., 2021). For the current paper, 2017 GERD data was used (the REF2014 midpoint) or the nearest available year.

## 4 Results

Additional graphs are available online at https://doi.org/10.6084/m9.figshare.23641209.

## 4.1 RQ1, H1: International collaboration overall

International collaboration as a binary variable in research quality regressions, after controlling for collaboration in general, associates statistically significantly with an increased probability of higher REF quality scores in all main panels (Figure 1). There is an international collaboration advantage in most (27 out of 34) UoAs, and it is statistically significant in just under half: 15 out of 34. Ignoring the results that are not statistically significant (e.g., UoA 31), international collaboration seems to have the greatest advantage in business and economics, and it is statistically significantly apparently a disadvantage only in the hybrid UoA 34: Communication, Cultural and Media Studies, Library and Information Management. The control variable for collaboration increases the odds of a higher quality score in 27 out of 34 UoAs and all four panels. The exceptions are UoAs 16, 17, 22, 28, 30, 31, and 32. The results do not support H1 (the advantage biggest in the social sciences and engineering) in that the collaboration advantage is negative for the large engineering UoA 12 and, although the two strongest statistically significant international advantages are for social sciences (UoAs 16 and 19) and two others are also high (UoAs 17 and 18), the rest do not have a statistically significant advantage. Moreover, the health and life sciences (Main Panel A) have a similar overall apparent benefit of internationalism to the social sciences (Main Panel B) (Figure 1).

There are clear differences between UoAs in the relative contributions of collaboration and internationalism. For example, collaboration is apparently a disadvantage for both UoA 17 Business and UoA 18 Economics but an international dimension to collaboration is a strong advantage. Conversely, although internationalism is never a statistically significant disadvantage, its role seems negligeable for some UoAs, including UoAs 2 Public Health, 3 Allied Health Professions, and 6 Agriculture, despite (any type of) collaboration being very important in them (Figure 1). Note that the two regression coefficients are not comparable because, for example, a coefficient of 2 for internationalism means that the odds ratio for a higher quality score is doubled for research with at least one non-UK author, whereas a coefficient of 2 for authorship means that the odds ratio for a higher quality score is doubled for an article with e=2.7 times more authors.

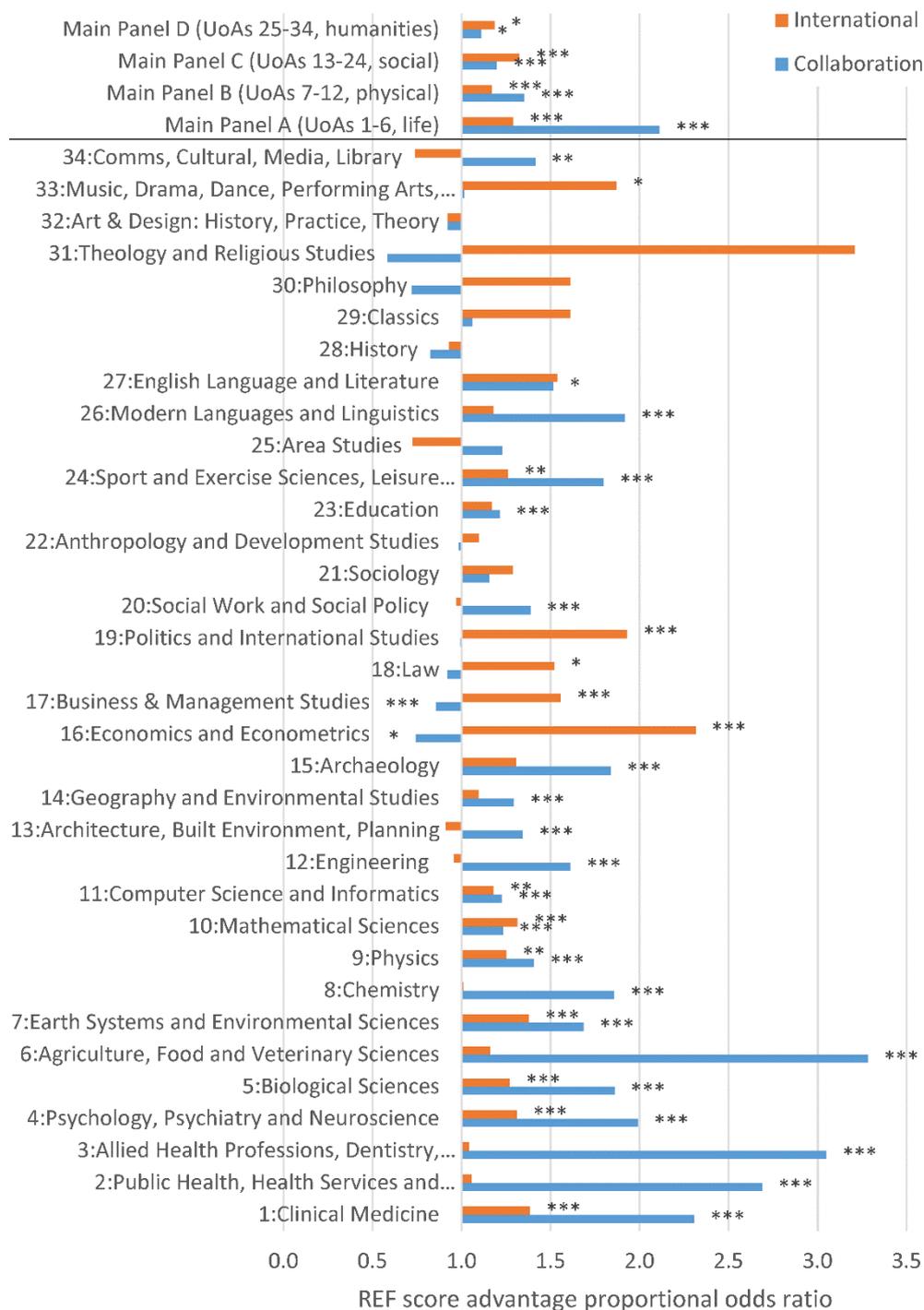

Figure 1. The average REF score advantage for UK journal articles having at least one international author (as a binary variable, orange bars), as calculated through 38 ordinal logistic regressions with log(authors) as a general collaboration control variable that ignores nationality (blue bars) (eqn. 1). A value of 1 for International indicates no advantage (or disadvantage) for international authorship compared to UK-only authors, controlling for authorship team size. A value of 1 for Collaboration indicates no advantage (or disadvantage) for articles with more authors, controlling for author team internationalism. Stars indicate statistical significance: *** p<0.001; ** p<0.01; * p<0.05.

## 4.2 RQ2, H1: International collaboration by country

The second ordinal regression model (eqn. 2) gives superior information to simple descriptive statistics about the average REF score advantage (see Appendix) because it considers the relative contributions of all countries to an article, as well as controlling for collaboration in general. The regression for Main Panel A confirms that author contributions from some countries increase the odds of a higher REF score but contributions from other countries decrease them (Figure 2). Note that whilst articles with an author from China tend to score above UK-only articles (see Appendix), there is little or no differences in the logistic regression (Figure 2), showing the importance of the regression approach. This could occur, for example, if collaborations with China tended to also include the USA. Note that the regression coefficients exaggerate the influence of collaboration because of fractional counting. For example, the coefficient of 3.9 for Israel means that the odds ratio for a higher quality research score increases 3.9 times if 100% Israeli authored, which is impossible since at least one author must be from the UK. More realistically, if a team is 50% from the UK and 50% from Israel then the odds ratio for a higher quality score would increase 3.9x0.5=1.95 times in comparison to a 100% UK-authored paper.

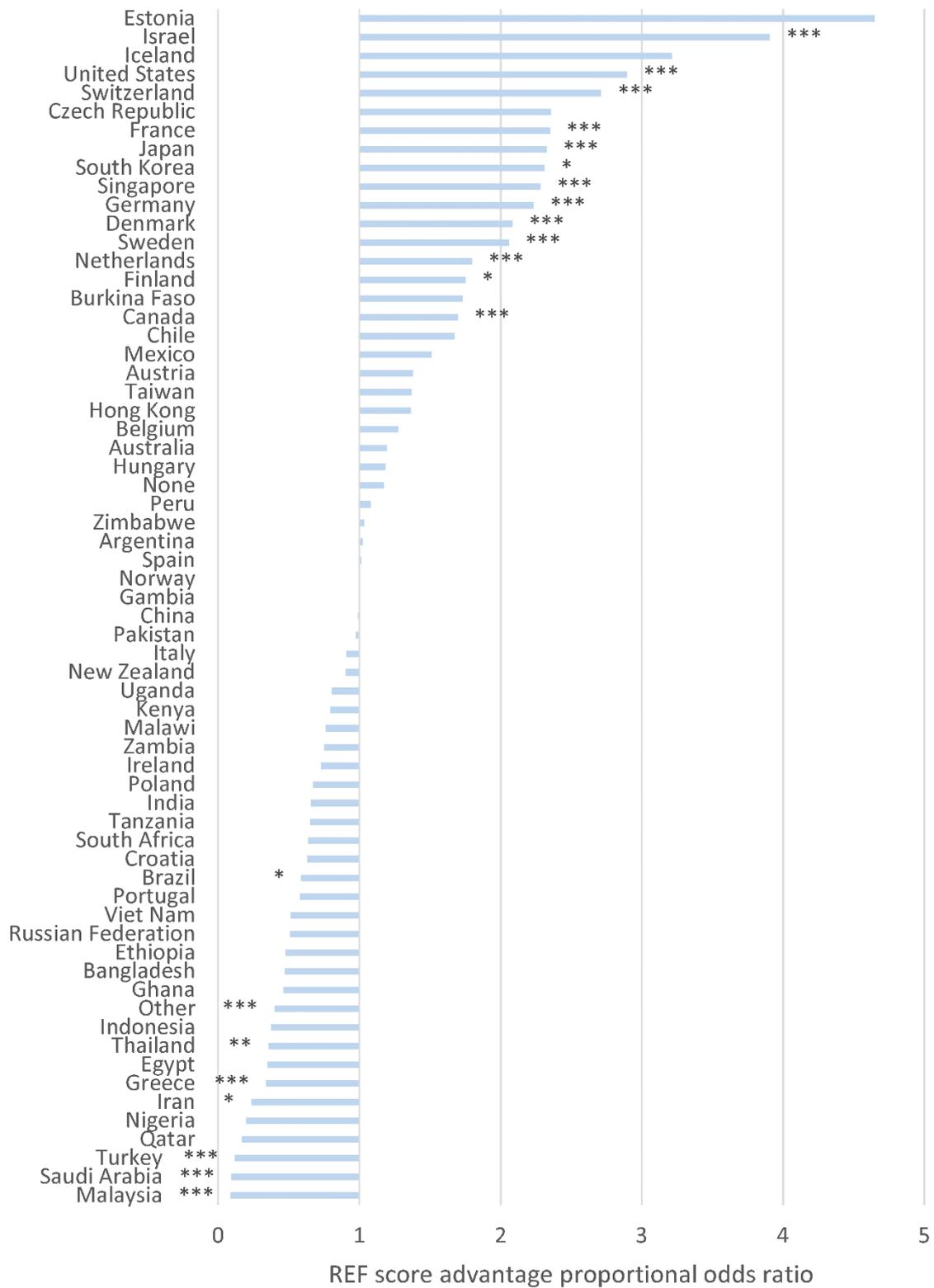

Figure 2. Odds ratios from an ordinal logistic regression for REF score for journal articles from Main Panel A, with log(authors) as a control variable (eqn. 2). Qualification: At least 10 journal articles using fractional author counting. Stars indicate statistical significance: *** p<0.001; ** p<0.01; * p<0.05. The logged authors odds ratio is 2.11. "None" means no country affiliation recorded in Scopus for an author.

For the mainly physical science and engineering articles of Main Panel B, the contributions of countries outside the UK are less likely to associate with a higher REF score overall than for Main Panel A. This is clearest from Figure A1 in the online supplement against the Appendix figure but is also evident by comparing Figure 3 to Figure 2. There are considerable overlaps in terms of countries with advantages in both Figures 3 and 2, or with disadvantages in both.

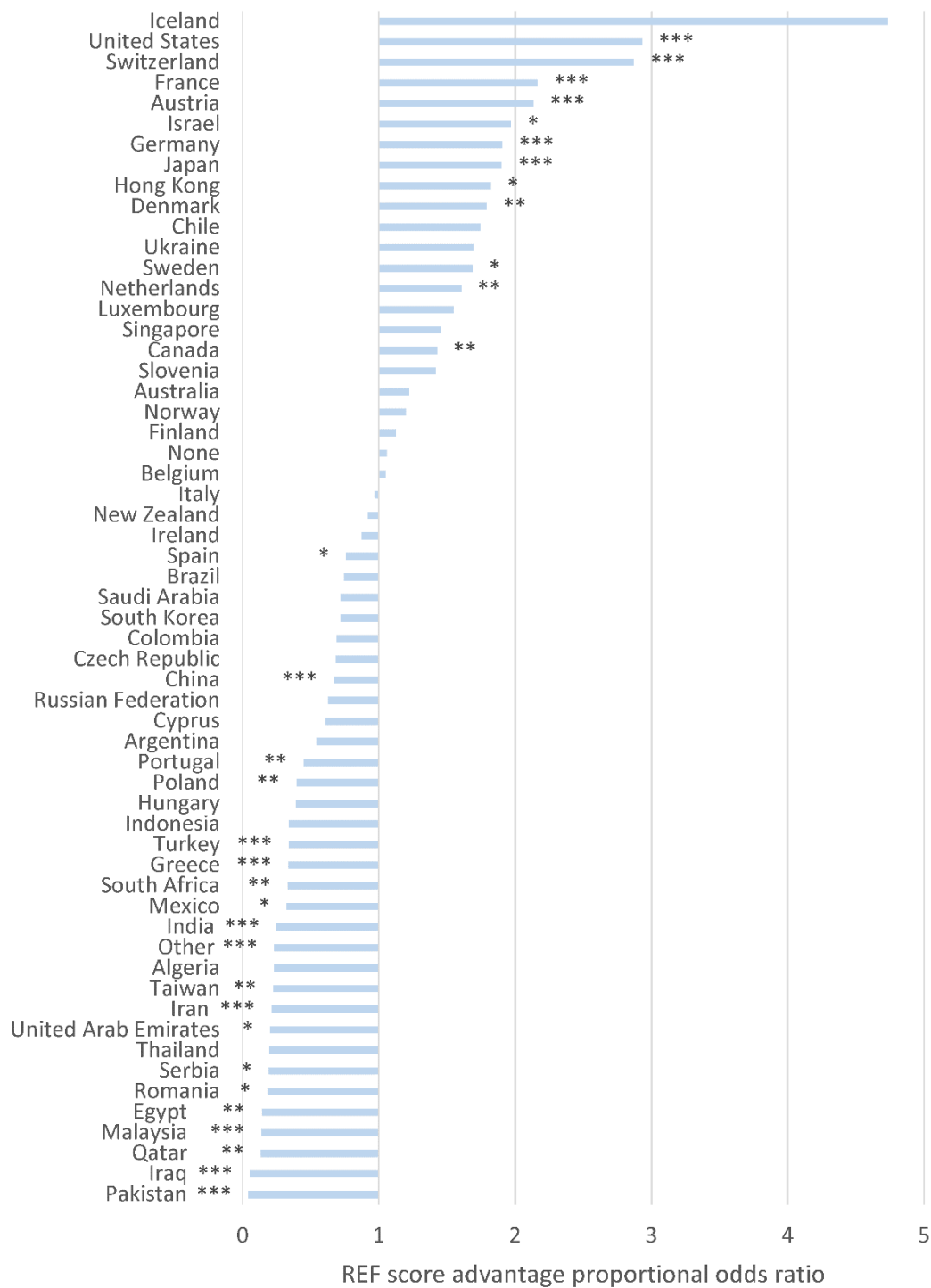

Figure 3. As Figure 2 for Main Panel B. The logged authors odds ratio is 1.31.

The predominantly social science research of Main Panel C shows advantages for many countries (Figure A2 in the online supplement, Figure 4). There are again considerable overlaps with Main Panel A and Main Panel B for countries with advantages in two of the three, or disadvantages in two of the three.

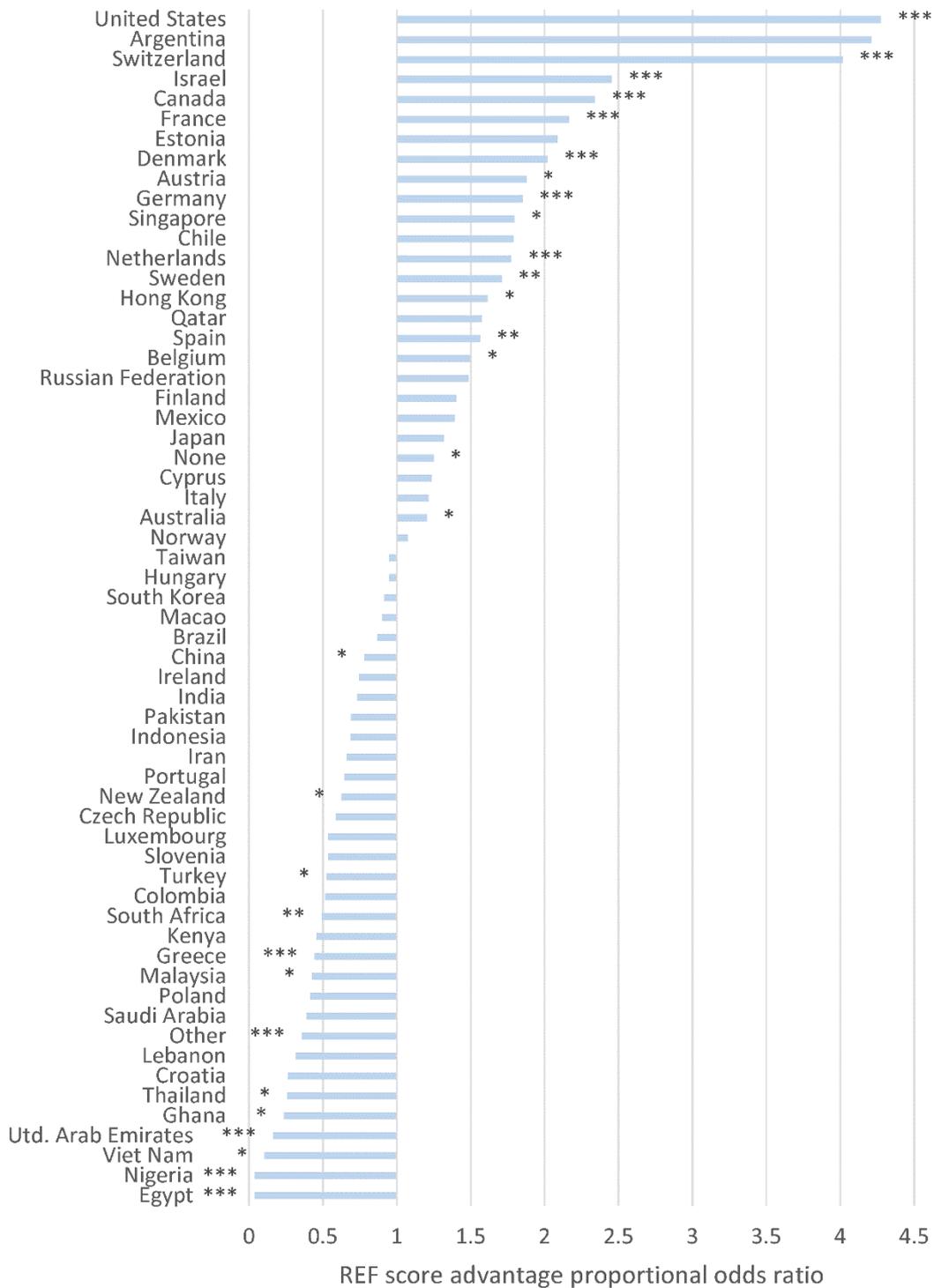

Figure 4. As Figure 3 for Main Panel C. The logged authors odds ratio is 1.24.

Main Panel D includes mainly arts and humanities research, with some social sciences fields too. It has relatively few journal articles because the UoAs tend to be small and researchers tend to submit books, book chapters, performances, or artworks instead. There is a weak tendency for non-UK contributions to benefit journal article quality (Figure A3 in the online supplement, Figure 5). The country with the strongest negative association, Spain, also has a negative association in Main Panel B, but a positive association in Main Panel C (and a weak non-significant association in Main Panel A). This confirms that there can be disciplinary differences in the apparent value of countries to international collaborations.

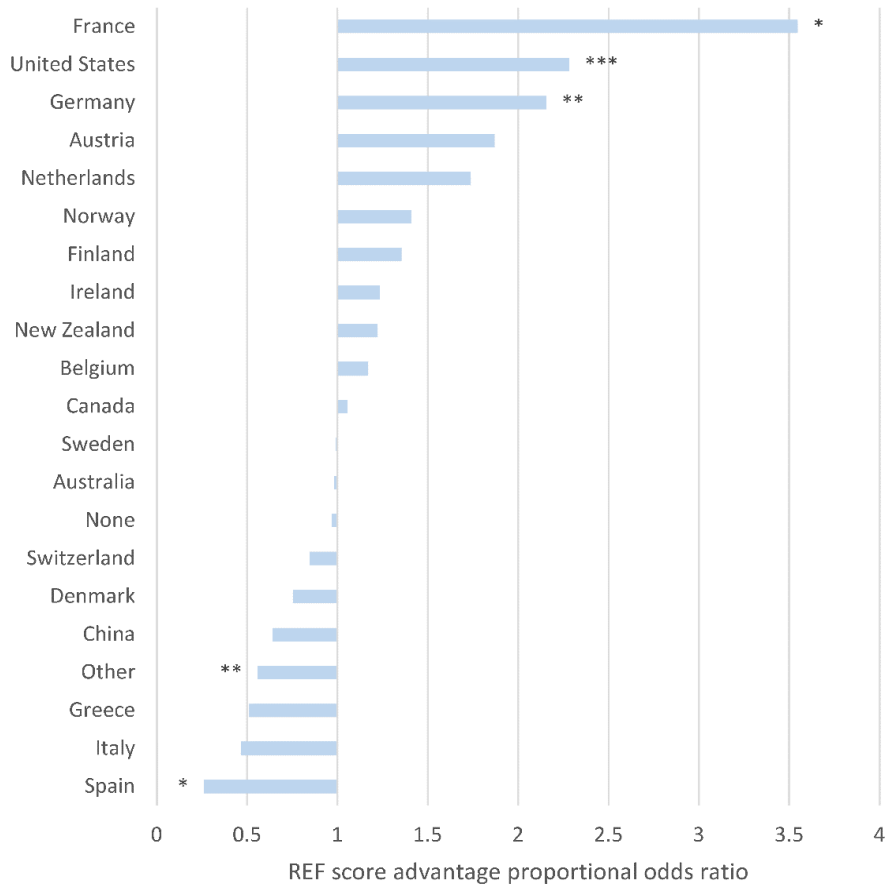

Figure 5. As Figure 3 for Main Panel D. The logged authors odds ratio is 1.14.

## 4.3 RQ3, H2, H3: Reasons for differing country collaboration advantages

This section investigates possible reasons for collaboration advantage differences between countries. There is a moderate positive correlation between the average citation impact of a country (MNLCS) and the ordinal regression coefficients for that country in Main Panels A, B, and C. The correlation is strong if only statistically significant coefficients are included (Table 1). This confirms that collaboration with countries independently producing higher citation impact research tends to give a greater likelihood of higher quality international collaborative research with the UK, supporting H2. Although citation rates can sometimes be misleading for small countries due to the dominance of important international collaborations or foreign-backed research institutes (Confraria et al., 2017), most of the countries analysed publish enough for this to be an unlikely alternative explanation.

Correlations with country regression coefficients are high for GERD (as a percentage of GDP) in all Main Panels, supporting H3. They are higher than for average national citation impact (MLNLCS), suggesting that national research spending may be more important for collaboration partners than highly cited research (Table 1). The correlations are highest for Main Panels A and B, plausibly suggesting that research expenditure is particularly important in health, life and physical science areas.

Table 1. Pearson correlations between significant regression coefficients (all regression coefficients) and (a) country MNLCS with fractional counting or (b) GERD as a percentage of GDP. The last two columns are the sample sizes.

| Panel | Country coefficients vs. MNLCS | Country coefficients vs. GERD as % of GDP | Countries | Countries with GERD |
|---|---|---|---|---|
| A life/health | 0.557 (0.334) | 0.893 (0.633) | 20 (62) | 20 (53) |
| B physical | 0.626 (0.459) | 0.833 (0.581) | 31 (55) | 31 (53) |
| C social | 0.617 (0.337) | 0.707 (0.431) | 27 (56) | 25 (50) |
| D humanities | -0.093 (0.004) | 0.591 (0.353) | 4 (19) | 4 (18) |

At the individual country level, in Main Panels A to C, the international collaboration advantage applies to all countries that exceed the UK's MNLCS (1.17) and many that don't (Figure 6; see supplementary materials for B, C and D). The same is true for Main Panel D (only four countries). The closest to an exception is Qatar in Main Panel B, which has an MNLCS close to the UK's but its collaborations with the UK tend to reduce the odds of a higher quality article, from the UK perspective. In broad terms, the results suggest that UK collaborations with other richer countries tend to be beneficial for the quality of research, irrespective of the citation impact of the collaborating country. There are exceptions to this rule, however, such as Spain for Main Panel B (a disadvantage) although Spanish collaboration is an advantage for Main Panel C.

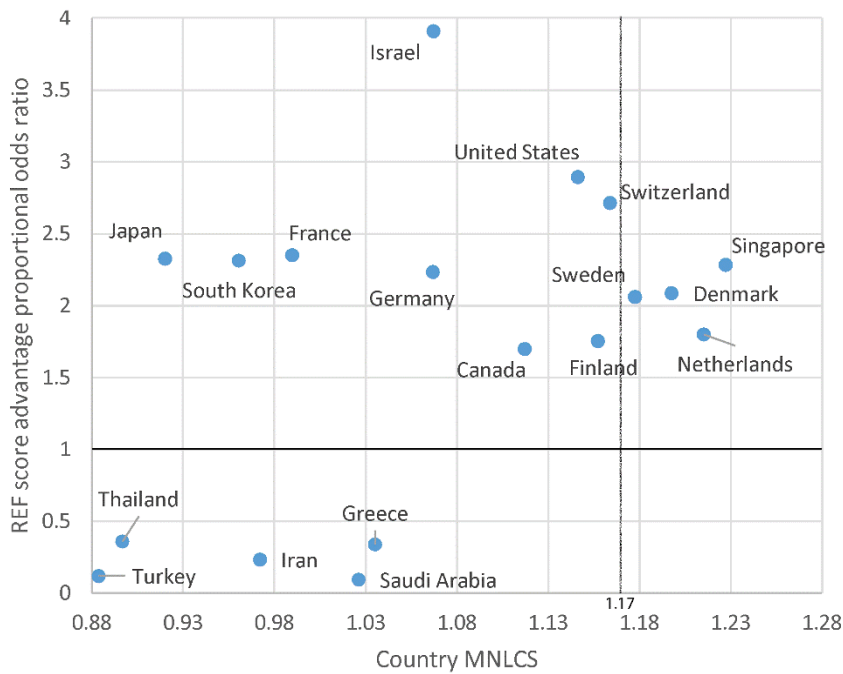

Figure 6. REF score advantage odds ratio against country MNLCS (fractional counting) for Main Panel A (mainly health and life sciences) UoAs combined, with only statistically significant country ratios shown. The UK MNLCS is 1.17

For Panel A overall (Figure 7) and to a large extent for Panel B and Panel C (see supplementary materials), the odds of higher quality research scores increase for countries with a higher GERD than the UK but decrease for countries with a lower GERD than the UK. As discussed above, there is no equivalent pattern for MNLCS (Figure 6) which supports H3 being stronger than H2: national research expenditure seems to be more important than national citation impact for the quality of collaborative research.

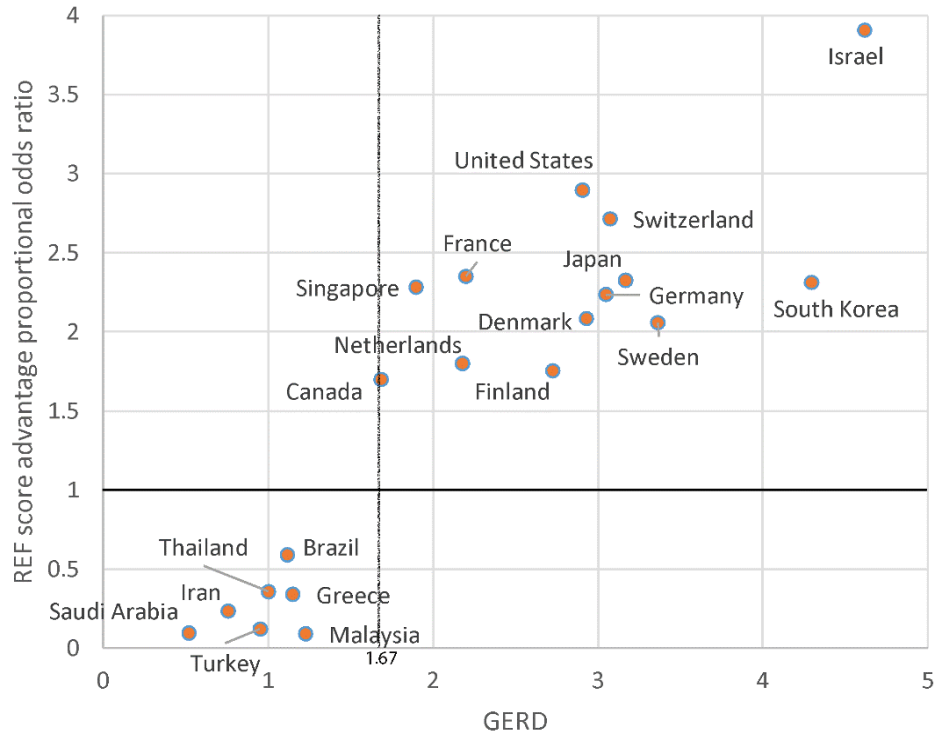

Figure 7. REF score advantage odds ratio against national research spending (GERD as a percentage of GDP) for Main Panel A (mainly health and life sciences) UoAs combined, with only statistically significant country ratios shown. The UK GERD is 1.67.

## 5   Discussion

The results are limited to the UK although they may apply to broadly similar countries in terms of the national average citation impact and research expenditure (GERD). The results are also limited to the self-selected UK REF submissions, which represent articles considered to be the authors' best and therefore underrepresent weaker articles. The results may therefore be different (stronger or weaker) for all articles. In UoA 8 Chemistry and UoA 9 Physics, researchers may have often written far more articles than they submitted. If they had co-authored many low-quality national or international articles, then these would be hidden from the REF due the selection process. The statistical trends identified will have many exceptions and may only concern a minority of the articles because of a confounding factor not included in the analysis.

Importantly, the results also concern research quality solely from a UK Global North perspective and very different patterns may have been found if Global South priorities had been used instead. This would especially effect perceptions of the value of UK collaborations with Global South countries.

The results do not prove a cause-and-effect relationship between international collaboration and research quality, for the reasons discussed above. For example, a researcher may use their best idea to attract funding for an international project rather than the international collaboration generating the good idea or otherwise stronger research. Similarly, better regarded UK researchers might find it easier to attract international partners for collaborative research. From a different perspective, since there are many funding programmes exclusively for international research, such as most from the EU framework programmes, international research may be better funded than national research, or may attract better participants because they were successful in funding competitions. In these cases, the international dimension would not improve the result directly but would be an effect of the underlying cause. There may also be intermediate factors that affect the results. For instance, perhaps it is easier to get an internationally collaborative article in a higher impact journal and this may have influenced the article's quality score in a few cases, even though assessors were explicitly told to ignore journal rankings and impact factors.

The strong positive relationship between international research collaboration and research quality contrasts with the very weak relationship found in the most relevant prior study, although it did not use fractional counting (Bornmann, 2017). The F1000Prime articles would probably fit mainly in UoA 1, which had a strong positive association with international collaboration. The difference may be due to the UK focus here, changes over time, the different model for accounting for the effect of international collaboration, or the different purposes of the quality scores used. The F1000Prime system used three ratings, 1: good; 2: very good; 3: exceptional and these were awarded by at least two peer reviewers, so this is a close parallel with the REF (four ratings, but 1* is rare, and usually two reviewers). Presumably the F1000Prime reviewers would be less uniformly senior than REF assessors and their judgments may have been less seriously considered and less norm referenced. F1000Prime articles are selected by the reviewers so may be more uniformly high quality than REF articles. This would reduce the statistical power of any tests. Thus, the main differences are probably the scope (author-selected UK articles vs. reviewer-selected international articles), the fractional counting model, and the quality of the reviewing processes. The stronger results compared to a previous REF study (HEFCE, 2015) are probably due to the fractional counting model and parsimonious independent variables.

The disciplinary differences found do not support the hypothesis based on prior research (rejecting H1) that international collaboration is particularly important for social sciences and engineering and instead finds that it is apparently most beneficial in the social sciences and the life and health sciences, although it is usually apparently beneficial in the physical sciences, arts and humanities. This may be due to the greater funding provided for typical health and life sciences research, such as for equipment, and consumables (and ethical approval procedures in health fields) (Huang & Huang, 2018), the benefits of multiple social contexts for social science research, or the need for international collaboration to address global sustainable development goals. The funding for international projects might be from large international competitive grants (e.g., from the European Union) or it might be indirect in the form of access to large international equipment, such as particle accelerators or space telescopes. The apparent lack of importance of international collaboration for UK engineering may reflect UK international engineering projects often having overseas development goals (e.g., proposing novel building materials for low resource environments) that were not valued by the assessors.

The finding that the outputs of collaborations with countries that with lower average citation impact (supporting H2) and lower research expenditure (supporting H3) tend to be of lower quality than similarly collaborative national UK research is worrying from a global equity perspective. There are multiple possible explanations. Global South research is less cited (Confraria et al., 2017) and may be lower quality than Global North research (from a Global North research evaluation perspective) due to lower English proficiency, less funding (Man et al., 2004), weaker research training, and less research support. This aligns with the evidence that Global South led research with Global North collaborators does not get a citation advantage (Chinchilla-Rodríguez et al., 2019). On this basis, a Global South co-author might tend to supply less expertise and resources than a Global North contributor, although they would surely contribute new perspectives. Collaborations with UK researchers might then have a partly translatory goal of helping Global South researchers to access a Global North audience (Martinez & Sá, 2020) and might receive less funding on this basis. The collaborative articles may thus be steps towards independence for the Global South researcher (at least towards addressing Global North research norms) in a way that the UK REF output evaluation system does not recognise, although it might in the research environment component of the REF. Development funding agencies are influenced by political considerations and often emphasise ambitious non-academic goals, which may undermine the academic scoring of such projects (Currie-Alder, 2015).

The above paragraph takes a Global North perspective on research quality. Alternatively, the UK REF might not recognise the goals, priorities, knowledge, and skills of Global South researchers and thus unfairly award lower quality scores to research that they led or participated in. A similar issue has been raised for Maori research in New Zealand (Roa et al., 2009). To give an extreme example, Dengue science might be considered much less important than cancer research in the UK, even though this mosquito-borne viral infection is an increasing threat in tropical and subtropical countries (Wellekens et al., 2022). Global South research might also sometimes have perspectives that some Global North evaluators might find unpalatable (Giwa, 2015; Openjuru et al., 2015) or inappropriate, such as criticism of colonial legacies or ongoing imperialism, or a greater focus on ethics and societal value.

In between the above two perspectives, it is also possible that REF evaluators have started to recognise and penalise exploitative helicopter science (Minasny et al., 2020; Schutte et al., 2023) and have downgraded research that they perceived to have a token Global South author without genuine Global South engagement. Thus, there are three plausible underlying reasons for H2 being supported.

# 6   Conclusion

Although not definitive and oriented on the UK, the results give the first large scale direct science-wide academic evidence that journal articles written by international teams tend to be higher quality (from a UK Global North perspective) than national research, even after accounting for team size. Whilst research quality is neither absolute nor universal, the judgements were made by experts that were not asked to consider internationality of the research team in any way. The internationalism advantage seems to apply to nearly all partner countries that invest at least as much in research (GERD as a percentage of GDP) as the UK, even those with low citation impact research. The overall international collaboration advantage for the UK is due to collaboration with higher research expenditure economies being more frequent than collaboration with other economies, since these tend to produce lower quality research, at least as judged by REF panel members.

From a pure research quality perspective, as judged by Global North standards, collaborations with lower research expenditure countries are not optimal for research quality, as judged by REF experts. This could be due to the research tending to be lower quality under any reasonable criteria. Alternatively, the results could be due to bias against these collaborations in the judgements made by the experts, such as for the scope of what is considered important (e.g., perhaps mainly Global North concerns). Although our data does not reveal the underlying cause, the most worrying possibility is that UK assessors are biased in that they undervalue research that is high quality from a Global South perspective. This should be investigated as a matter of urgency. Moreover, these collaborations seem likely to satisfy wider goals, such as creating international connections that may be useful for other purposes (Wagner, 2008) or trying to overcome the legacies of colonialism (especially for the UK) (Allpress et al., 2010). Even for the UK REF, researchers may be rewarded for achieving these wider goals through the environment and impact case study REF components. Moreover, these results are all relative, since all the collaborations analysed allowed a UK academic to produce (what they or their institution considered) their best research since it was selected in the first place for the REF.

International collaboration, compared to general collaboration, is advantageous in most but not all fields and seems particularly useful in the health, life, and social sciences. Engineering is a surprising exception in terms of a field that apparently does not benefit from international collaboration, at least from a UK perspective.

In terms of research policy implications, the results suggest that funders and research managers should continue to generally encourage international collaborative research because the journal articles produced tend to be higher quality. Policy makers should also be aware that international collaboration may be disadvantageous in some fields, however (Figure 1) and have little effect in others, at least from the UK perspective and with the REF definition of quality. In this context, it seems sensible to be judicious in selecting fields to target for international research funding and to allow national research to blossom in fields where it is more likely to be excellent.

The online supplementary materials are at https://doi.org/10.6084/m9.figshare.23641209.

## 7 Acknowledgement

This study was funded by Research England, Scottish Funding Council, Higher Education Funding Council for Wales, and Department for the Economy, Northern Ireland as part of the Future Research Assessment Programme (https://www.jisc.ac.uk/future-research-assessment-programme). The funders had no role in the design or execution of this study. The content is solely the responsibility of the authors and does not necessarily represent the official views of the funders.

# 9 Appendix

From a descriptive statistics perspective without controlling for collaboration or using regression, average quality scores by country (fractional counting) gives an overview of international quality differences in UK co-authored outputs. For Main Panel A (UoAs 1-6), which mainly covers the health and life sciences, co-authorship with many countries associates with higher average research quality scores than domestic UK authorship (Figure 8). For example, articles with an author from Iceland, on average, scored 0.46 higher than articles from only UK authors and articles from Malaysia, on average scored 0.42 lower, using fractional counting. Most countries have an apparent score advantage compared to the UK, partly because of the collaboration advantage of international authorship, which is not controlled for in Figure 8.

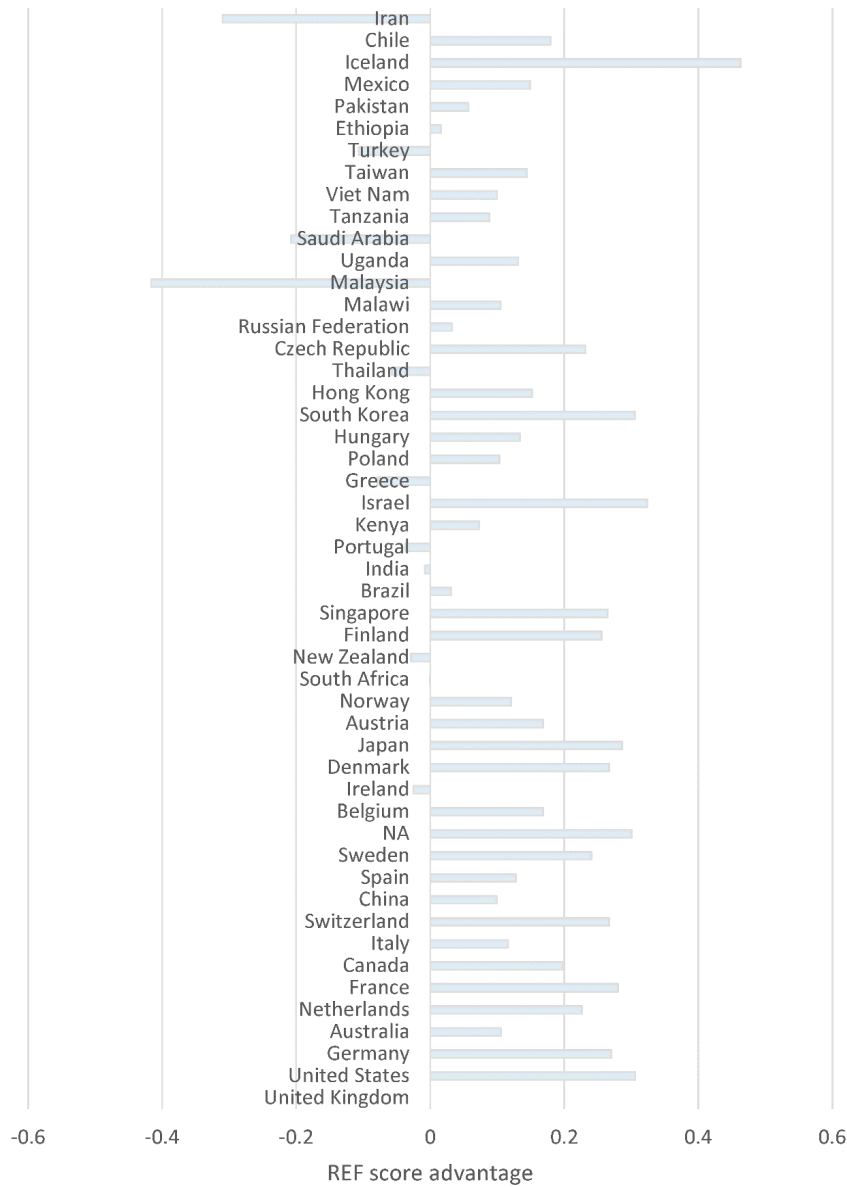

Figure 8. The weighted mean average score advantage for articles with an author from the specified country for articles submitted to Main Panel A (mainly health and life sciences). The reference score is the UK average. Qualification: the 50 countries co-authoring the most Main Panel A journal articles, using fractional author counting. NA: no country assigned in Scopus. Countries are ordered by the number of co-authored articles in the dataset.